\documentclass[twocolumn]{article}

\usepackage{listings, textcomp}
\usepackage{graphicx,xcolor}
\RequirePackage[colorlinks=true, allcolors=blue]{hyperref}

\usepackage{caption}
\usepackage{titlesec}
\titleformat*{\section}{\large\bfseries}
\title{Array Programming with NumPy}

\RequirePackage{authblk}
\author[1]{Charles R. Harris}
\author[2,3,4,*]{K. Jarrod Millman}
\author[5,2,4,*]{St\'efan J. van der Walt}
\author[6,*]{Ralf Gommers}
\author[7]{Pauli Virtanen}
\author[8]{David Cournapeau}
\author[9]{Eric Wieser}
\author[10]{Julian Taylor}
\author[4]{Sebastian Berg}
\author[11]{Nathaniel J. Smith}
\author[12]{Robert Kern}
\author[4]{Matti Picus}
\author[13]{Stephan Hoyer}
\author[14]{Marten H. van Kerkwijk}
\author[2,15]{Matthew Brett}
\author[16]{Allan Haldane}
\author[17]{Jaime Fern\'andez del R\'io}
\author[18,19]{Mark Wiebe}
\author[6,20,21]{Pearu Peterson}
\author[22,23]{Pierre G\'erard-Marchant}
\author[24]{Kevin Sheppard}
\author[25]{Tyler Reddy}
\author[4]{Warren Weckesser}
\author[6]{Hameer Abbasi}
\author[26]{Christoph Gohlke}
\author[6]{Travis E. Oliphant}
\affil[1]{Independent Researcher, Logan, Utah, USA}
\affil[2]{Brain Imaging Center, University of California, Berkeley, Berkeley, CA, USA}
\affil[3]{Division of Biostatistics, University of California, Berkeley, Berkeley, CA, USA}
\affil[4]{Berkeley Institute for Data Science, University of California, Berkeley, Berkeley, CA, USA}
\affil[5]{Applied Mathematics, Stellenbosch University, Stellenbosch, South Africa}
\affil[6]{Quansight LLC, Austin, TX, USA}
\affil[7]{Department of Physics and Nanoscience Center, University of Jyv\"askyl\"a, Jyv\"askyl\"a, Finland}
\affil[8]{Mercari JP, Tokyo, Japan}
\affil[9]{Department of Engineering, University of Cambridge, Cambridge, UK}
\affil[10]{Independent Researcher, Karlsruhe, Germany}
\affil[11]{Independent Researcher, Berkeley, CA, USA}
\affil[12]{Enthought, Inc., Austin, TX, USA}
\affil[13]{Google Research, Mountain View, CA, USA}
\affil[14]{Department of Astronomy \& Astrophysics, University of Toronto, Toronto, ON, Canada}
\affil[15]{School of Psychology, University of Birmingham, Edgbaston, Birmigham, UK}
\affil[16]{Department of Physics, Temple University, Philadelphia, PA, USA}
\affil[17]{Google, Zurich, Switzerland}
\affil[18]{Department of Physics and Astronomy, The University of British Columbia, Vancouver, BC, Canada}
\affil[19]{Amazon, Seattle, Washington, USA}
\affil[20]{Independent Researcher, Saue, Estonia}
\affil[21]{Department of Mechanics and Applied Mathematics, Institute of Cybernetics at Tallinn Technical University, Tallinn, Estonia}
\affil[22]{Department of Biological and Agricultural Engineering, University of Georgia, Athens, GA}
\affil[23]{France-IX Services, Paris, France}
\affil[24]{Department of Economics, University of Oxford, Oxford, UK}
\affil[25]{CCS-7, Los Alamos National Laboratory, Los Alamos, NM, USA}
\affil[26]{Laboratory for Fluorescence Dynamics, Biomedical Engineering Department, University of California, Irvine, Irvine, CA, USA}
\affil[*]{millman@berkeley.edu, stefanv@berkeley.edu, ralf.gommers@gmail.com}
 
\begin{document}

\flushbottom
\maketitle
\thispagestyle{empty}

\twocolumn[
  \begin{@twocolumnfalse}
    \begin{abstract}
     Array programming provides a powerful, compact, expressive syntax for accessing,
     manipulating, and operating on data in vectors, matrices, and
     higher-dimensional arrays \cite{iverson1980notation}.
     NumPy is the primary array programming library for the Python language
     \cite{dubois2007guest,oliphant2007python,millman2011python,perez2011python}.
     It plays an essential role in research analysis pipelines in fields as
     diverse as physics, chemistry, astronomy, geoscience, biology, psychology,
     material science, engineering, finance, and economics.
     For example, in astronomy, NumPy was an important part of the software stack used
     in the discovery of gravitational waves \cite{abbott2016observation}
     and the first imaging of a black hole \cite{eht-imaging}.
     Here we show how a few fundamental array concepts lead to a simple and
     powerful programming paradigm for organizing, exploring, and analyzing
     scientific data.
     NumPy is the foundation upon which the entire scientific Python
     universe is constructed. It is so pervasive that several projects,
     targeting audiences with specialized needs, have developed their own
     NumPy-like interfaces and array objects. Because of its central position in the
     ecosystem, NumPy increasingly plays the role of an interoperability layer
     between these new array computation libraries.
     \end{abstract}
    \vspace{1cm}
  \end{@twocolumnfalse}
]

\begin{figure*}[h]
  \centering
  \includegraphics[width=\textwidth]{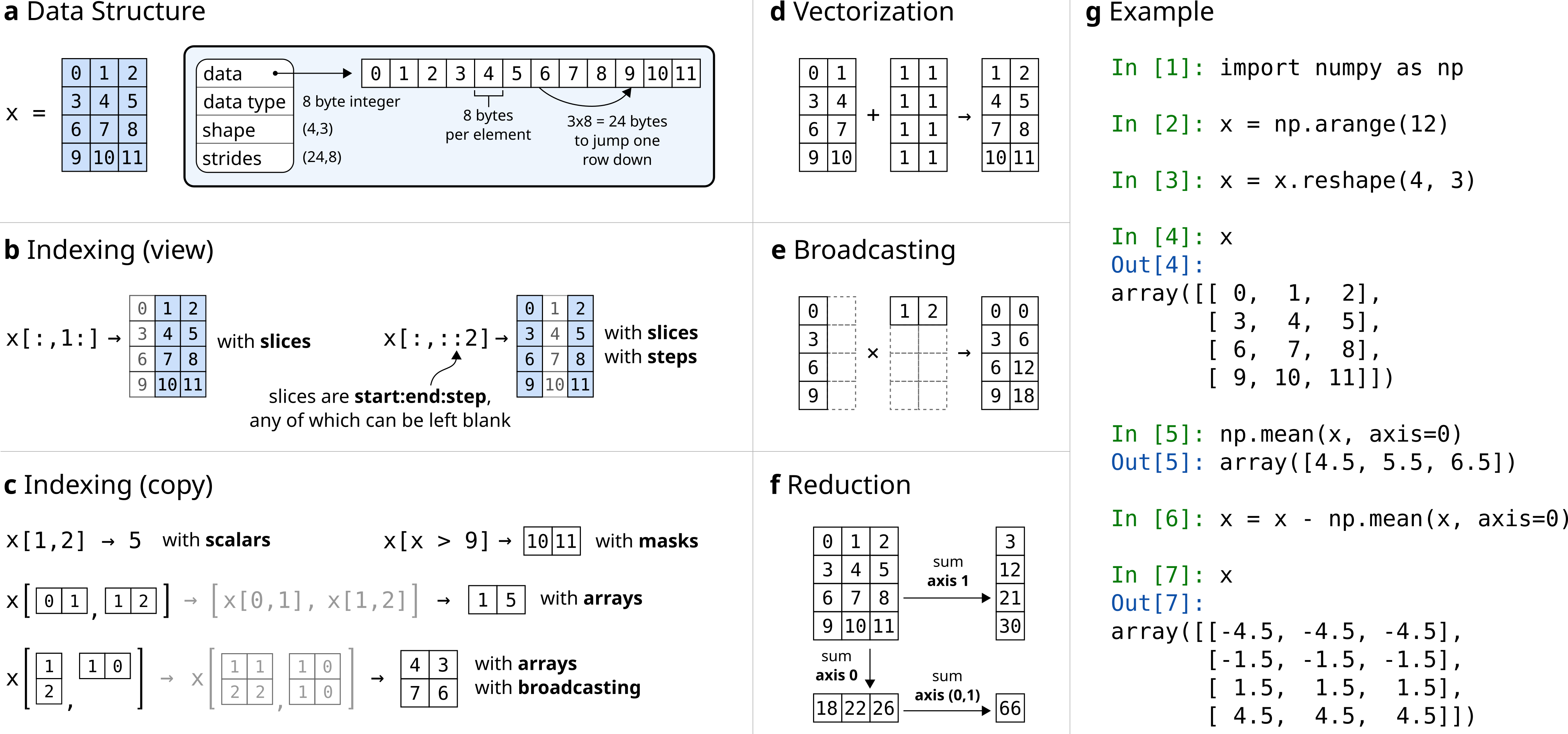}
  \caption{\textbf{The NumPy array incorporates several fundamental array concepts.}
    \textbf{a,} The NumPy array data structure and its associated metadata fields.
    \textbf{b,} Indexing an array with slices and steps. These
  operations return a \emph{view} of the original data.
    \textbf{c,} Indexing an array with masks, scalar coordinates, or
  other arrays, so that it returns a copy of the original data. In the
  bottom example, an array is indexed with other arrays; this broadcasts the indexing arguments before performing the lookup.
    \textbf{d,} Vectorization efficiently applies operations to groups
  of elements.
    \textbf{e,} Broadcasting in the multiplication of two-dimensional arrays.
    \textbf{f,} Reduction operations act along one or more axes. In this
    example, an array is summed along select axes to produce a vector, or along two axes consecutively to
    produce a scalar.
    \textbf{g,} Example NumPy code, illustrating some of these concepts.
   }
  \label{fig:array-concepts}
\end{figure*}

Two Python array packages existed before NumPy.
The Numeric package began in the mid-1990s and provided an array object and array-aware functions
in Python, written in C, and linking to standard fast implementations of linear
algebra \cite{dubois1996numerical,Numericmanual}.
One of its earliest uses was to steer C++ applications for
inertial confinement fusion research at
Lawrence Livermore National Laboratory \cite{yang1997steering}.
To handle large astronomical images coming from the Hubble Space Telescope,
a reimplementation of Numeric, called Numarray, added
support for structured arrays, flexible indexing, memory mapping, byte-order variants,
more efficient memory use, flexible IEEE error handling capabilities, and
better type casting rules \cite{greenfield2003numarray}.
While Numarray was highly compatible with Numeric, the two packages had enough
differences that it divided the community, until 2005, when NumPy emerged as a
``best of both worlds'' unification \cite{oliphant2006guide}---combining
Numarray's features with Numeric's performance on small arrays and its rich C
\emph{Application Programming Interface} (API).

Now, fifteen years later, NumPy underpins almost every Python library that does scientific or
numerical computation including SciPy \cite{virtanen2020scipy},
Matplotlib \cite{hunter2007matplotlib}, pandas \cite{mckinney-proc-scipy-2010},
scikit-learn \cite{pedregosa2011scikit}, and
scikit-image \cite{vanderwalt2014scikit}.
It is a community-developed, open-source library, which provides a
multidimensional Python array object along with array-aware functions
that operate on it.
Because of its inherent simplicity, the NumPy array is
the \emph{de facto} exchange format for array data in Python.

NumPy operates on in-memory arrays using the CPU. To utilize modern,
specialized storage and hardware, there has been a recent
proliferation of Python array packages. Unlike with the Numarray and
Numeric divide, it is now much harder for these new libraries to
fracture the user community---given how much work already builds
on top of NumPy.  However, to provide the ecosystem with access to
new and exploratory technologies, NumPy is transitioning into a
central coordinating mechanism that specifies a well-defined array
programming API and
dispatches it, as appropriate, to specialized array implementations.

\section*{NumPy arrays}

The NumPy array is a data structure that efficiently stores and accesses
multidimensional arrays \cite{vanderwalt2011numpy}, also known as tensors, and
enables a wide variety of scientific computation.
It consists of a pointer to memory, along with metadata used to interpret the
data stored there, notably {\em data type}, {\em shape}, and {\em strides}
(Fig.~\ref{fig:array-concepts}a).

The \emph{data type} describes the nature of elements stored in an array.
An array has a single data type, and each array element occupies the same
number of bytes in memory.
Examples of data types include real and complex numbers (of lower and higher
precision), strings, timestamps, and pointers to Python objects.

The \emph{shape} of an array determines the number of elements along each axis,
and the number of axes is the array's dimensionality.
For example, a vector of numbers can be stored as a one-dimensional array of
shape $N$, while color videos are four-dimensional arrays of shape
$(T, M, N, 3)$.

\emph{Strides} are necessary to interpret computer memory, which stores elements
linearly, as multidimensional arrays.
It describes the number of bytes to move forward in memory to jump from row to
row, column to column, and so forth.
Consider, for example, a 2-D array of floating-point numbers with shape
$(4, 3)$, where each element occupies 8 bytes in memory.
To move between consecutive columns, we need to jump forward 8 bytes in memory,
and to access the next row $3 \times 8 = 24$ bytes.
The strides of that array are therefore $(24, 8)$.  NumPy can
store arrays in either C or Fortran memory order, iterating
first over either rows or columns.  This allows external libraries
written in those languages to access NumPy array data in memory directly.

Users interact with NumPy arrays using {\em indexing} (to access
subarrays or individual elements), {\em operators} (e.g., $+$, $-$, $\times$
for vectorized operations and $@$ for matrix multiplication), as well as {\em array-aware functions};
together, these provide an easily readable, expressive, high-level API for
array programming, while NumPy
deals with the underlying mechanics of making operations fast.

\emph{Indexing} an array returns single elements, subarrays, or elements that satisfy
a specific condition (Fig.~\ref{fig:array-concepts}b).
Arrays can even be indexed using other arrays (Fig.~\ref{fig:array-concepts}c).
Wherever possible, indexing that retrieves a subarray returns a {\em view} on
the original array, such that data is shared between the two arrays.
This provides a powerful way to operate on subsets of array data while
limiting memory usage.

To complement the array syntax, NumPy includes functions that perform
\emph{vectorized} calculations on arrays, including arithmetic, statistics, and
trigonometry (Fig.~\ref{fig:array-concepts}d).
Vectorization---operating on whole arrays rather than their individual
elements---is essential to array programming.
This means that operations that would take many tens of lines to express in
languages such as C can often be implemented as a single, clear Python
expression.
This results in concise code and frees users to focus on the details of
their analysis, while NumPy handles looping over array elements near-optimally,
taking into consideration, for example, strides, to best utilize the
computer's fast cache memory.

When performing a vectorized operation (such as addition) on two arrays with
the same shape, it is clear what should happen.
Through \emph{broadcasting}, NumPy allows the dimensions to differ, while
still producing results that appeal to intuition.
A trivial example is the addition of a scalar value to an array, but broadcasting also
generalizes to more complex examples such as scaling each column of an array
or generating a grid of coordinates.
In broadcasting, one or both arrays are virtually duplicated (that is, without
copying any data in memory), so that the shapes of the operands match
(Fig.~\ref{fig:array-concepts}d).
Broadcasting is also applied when an array is indexed using arrays of
indices (Fig.~\ref{fig:array-concepts}c).

Other array-aware functions, such as \texttt{sum}, \texttt{mean}, and \texttt{maximum}, perform
element-by-element \emph{reductions}, aggregating results across one,
multiple, or all axes of a single array.
For example, summing an $n$-dimensional array over $d$ axes results in a
$(n-d)$-dimensional array (Fig.~\ref{fig:array-concepts}f).

NumPy also includes array-aware functions for creating, reshaping, concatenating, and padding
arrays; searching, sorting, and counting data; and reading and writing files.
It provides extensive support for generating pseudorandom numbers,
includes an assortment of probability distributions, and
performs accelerated linear algebra, utilizing one of several backends
such as OpenBLAS \cite{wang2013augem,xianyi2012model} or Intel MKL optimized
for the CPUs at hand.

Altogether, the combination of a simple in-memory array
representation, a syntax that closely mimics mathematics, and a
variety of array-aware utility functions forms a productive and
powerfully expressive array programming language.

\begin{figure}
  \centering
  \includegraphics[width=.5\textwidth]{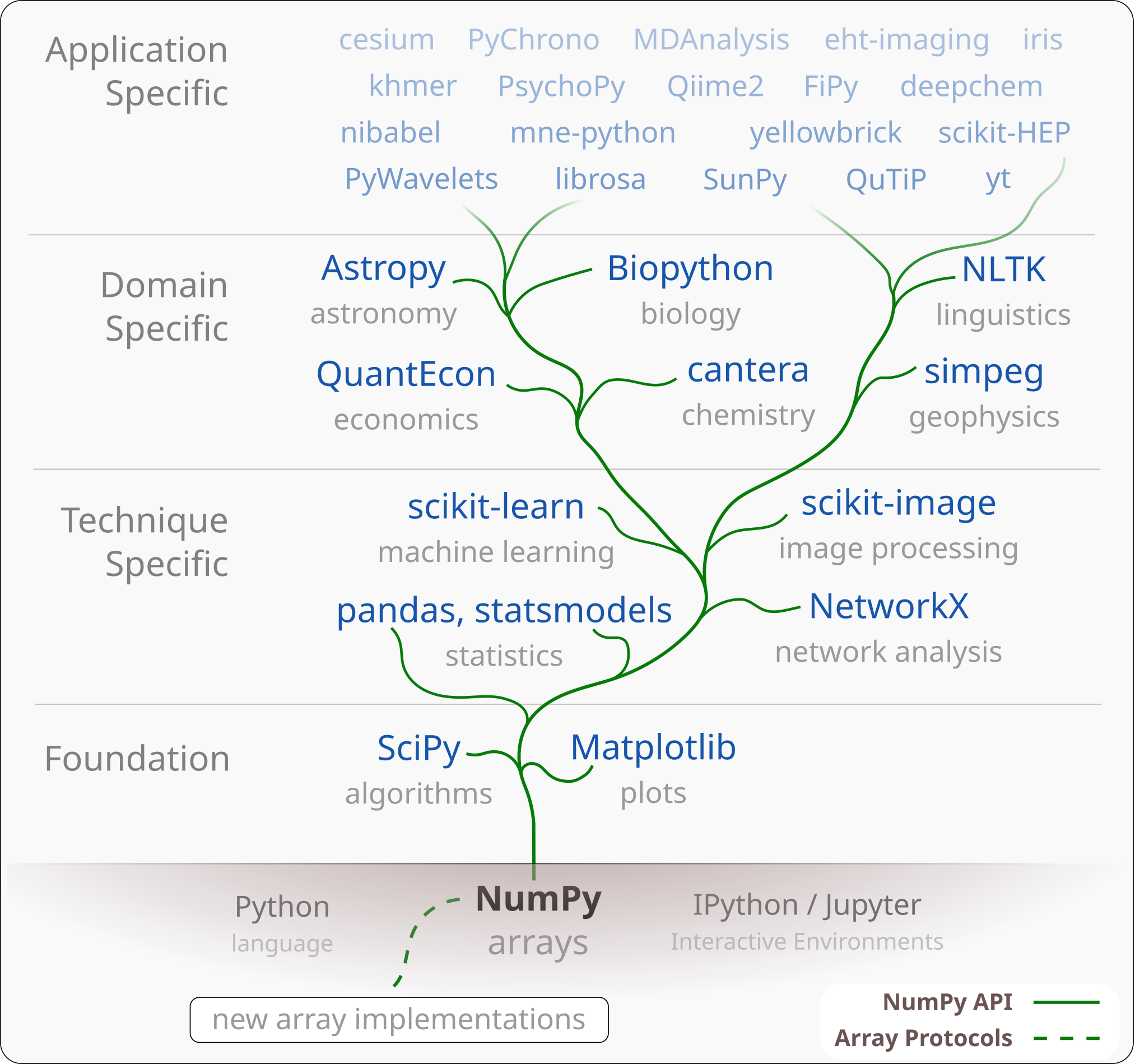}
  \caption{\textbf{NumPy is the base of the scientific Python ecosystem.}
   Essential libraries and projects that depend on NumPy's API gain access to new array
   implementations that support NumPy's array protocols (Fig.~\ref{fig:array-protocol}).
  }
  \label{fig:ecosystem}
\end{figure}

\section*{Scientific Python ecosystem}

Python is an open-source, general-purpose, interpreted programming language
well-suited to standard programming tasks such as cleaning data,
interacting with web resources, and parsing text.
Adding fast array operations and linear algebra allows scientists to do all
their work within a single language---and one that has the advantage of
being famously easy to learn and teach, as witnessed by its adoption
as a primary learning language in many universities.

Even though NumPy is not part of Python's standard library,
it benefits from a good relationship with the Python developers.
Over the years, the Python language has added new features and
special syntax so that NumPy would have a more succinct and 
easier to read array notation.
Since it is not part of the standard library, NumPy is able to
dictate its own release policies and development patterns.

SciPy and Matplotlib are tightly coupled with NumPy---in terms of
history, development, and use.
SciPy provides fundamental algorithms for scientific computing,
including mathematical, scientific, and engineering routines.
Matplotlib generates publication-ready figures and visualizations.
The combination of NumPy, SciPy, and Matplotlib, together with
an advanced interactive environment like IPython \cite{perez2007ipython},
or Jupyter \cite{Kluyver:2016aa}, provides a solid foundation for array
programming in Python.
The scientific Python ecosystem (Fig.~\ref{fig:ecosystem}) builds on top of
this foundation to provide several, widely used \emph{technique specific}
libraries \cite{pedregosa2011scikit,vanderwalt2014scikit,SciPyProceedings_11},
that in turn underlay numerous \emph{domain specific} projects
\cite{astropy:2013,astropy:2018,cock2009biopython,millman2007analysis,sunpy2015,2018EGUGA..2012146H}.
NumPy, at the base of the ecosystem of array-aware libraries,
sets documentation standards, provides array testing infrastructure,
and adds build support for Fortran and other compilers.

Many research groups have designed large,
complex scientific libraries, which add \emph{application specific} functionality
to the ecosystem.
For example, the \texttt{eht-imaging} library \cite{chael2019ehtim} developed by
the Event Horizon Telescope collaboration for radio interferometry imaging,
analysis, and simulation, relies on many lower-level components of the scientific Python
ecosystem.
NumPy arrays are used to store and manipulate numerical data at every step
in the processing chain: from raw data through calibration and image
reconstruction.
SciPy supplies tools for general image processing tasks such as
filtering and image alignment, while scikit-image, an image processing
library that extends SciPy, provides higher-level functionality such as
edge filters and Hough transforms.
The \texttt{scipy.optimize} module performs mathematical optimization.
NetworkX \cite{SciPyProceedings_11}, a package for complex
network analysis, is used to verify image comparison consistency.
Astropy \cite{astropy:2013, astropy:2018} handles standard
astronomical file formats and computes time/coordinate transformations.
Matplotlib is used to visualize data and to generate the final image of the black hole.

The interactive environment created by the array programming
foundation along with the surrounding ecosystem of tools---inside of
IPython or Jupyter---is ideally suited to exploratory data analysis.
Users fluidly inspect, manipulate, and visualize their data, and
rapidly iterate to refine programming statements. These statements are
then stitched together into imperative or functional programs, or
notebooks containing both computation and narrative.
Scientific computing beyond exploratory work is often done in a text editor
or an integrated development environment (IDEs) such as Spyder.
This rich and productive environment has made Python popular
for scientific research.

To complement this facility for exploratory work and rapid
prototyping, NumPy has developed a culture of
employing time-tested software engineering practices to improve collaboration and
reduce error \cite{millman2014developing}.  This culture is not only
adopted by leaders in the project but also enthusiastically taught to
newcomers. The NumPy team was early in adopting distributed revision
control and code review to improve collaboration on code, and
continuous testing that runs an extensive battery of automated tests for
every proposed change to NumPy.  The project also has comprehensive,
high-quality documentation, integrated with the source
code \cite{vanderwalt2008scipy,harrington2008scipy,harrington2009scipy}.

This culture of using best practices for producing reliable scientific software
has been adopted by the ecosystem of libraries that build on NumPy.
For example, in a recent award given by the Royal Astronomical Society to
Astropy, they state:
\begin{quotation}
\noindent\emph{The Astropy Project has provided hundreds of junior scientists
with experience in professional-standard software development practices
including use of version control, unit testing, code review and issue tracking
procedures. This is a vital skill set for modern researchers that is often
missing from formal university education in physics or astronomy.}
\end{quotation}
Community members explicitly work to address this lack of formal education
through courses and workshops
\cite{wilson-software-carpentry,hannay-scientific-software-survey,millman2018teaching}.

The recent rapid growth of data science, machine learning, and
artificial intelligence has further and dramatically boosted the usage of
scientific Python.  Examples of its significant application, such as the
\texttt{eht-imaging} library, now exist in almost every discipline in the natural and social
sciences.  These tools have become \emph{the primary}
software environment in many fields.  NumPy and its ecosystem are commonly
taught in university courses, boot camps, and summer schools, and are
at the focus of community conferences and workshops worldwide.

NumPy and its API have become truly ubiquitous.

\section*{Array proliferation and interoperability}

\begin{figure*}
  \centering
  \includegraphics[width=\textwidth]{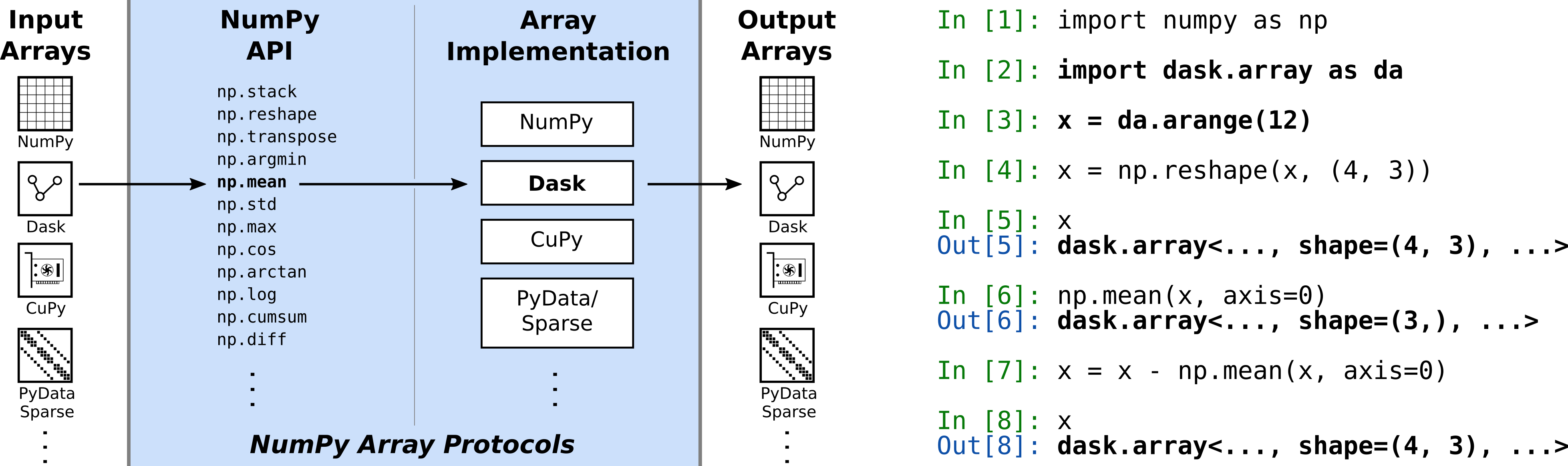}

  \caption{\textbf{NumPy's API and array protocols expose new arrays to the ecosystem.}
     In this example, NumPy's \texttt{mean} function is called on a Dask
     array.  The call succeeds by dispatching to the appropriate library implementation
     (i.e., Dask in this case) and results in a new Dask array.  Compare this
     code to the example code in Fig.~\ref{fig:array-concepts}g.
  }\label{fig:array-protocol}

\end{figure*}

NumPy provides in-memory, multidimensional, homogeneously typed
(i.e., single pointer and strided) arrays on CPUs.  It runs on machines
ranging from embedded devices to the world's largest supercomputers,
with performance approaching that of compiled languages.
For most its existence, NumPy addressed the vast majority of
array computation use cases.

However, scientific data sets now routinely exceed the memory capacity of a single machine and may
be stored on multiple machines or in the cloud.
In addition, the recent need to accelerate deep learning and artificial intelligence applications
has led to the emergence of specialized accelerator hardware,
including graphics processing units (GPUs), tensor processing units (TPUs),
and field-programmable gate arrays (FPGAs).
Due to its in-memory data model, NumPy is currently unable to
utilize such storage and specialized hardware directly.  However, both
distributed data and the parallel execution of GPUs, TPUs, and FPGAs map well
to the \emph{paradigm} of array programming: a gap, therefore, existed between
available modern hardware architectures and the tools necessary to
leverage their computational power.

The community's efforts to fill this gap led to a
proliferation of new array implementations. For example, each deep learning framework created
its own arrays; PyTorch \cite{NEURIPS2019_9015},
Tensorflow \cite{abadi2016tensorflow}, Apache MXNet \cite{chen2015mxnet},
and JAX arrays all have the
capability to run on CPUs and GPUs, in a distributed fashion, utilizing lazy evaluation
to allow for additional performance optimizations.  SciPy and PyData/Sparse both
provide sparse arrays---which typically contain few non-zero values and store
only those in memory for efficiency.
In addition, there are projects that build on top of NumPy arrays as a data
container and \textit{extend} its capabilities.  Distributed arrays are
made possible that way by Dask, and labeled arrays---referring to dimensions of
an array by name rather than by index for clarity, compare \texttt{x[:,~1]} vs.
\texttt{x.loc[:,~'time']}---by xarray \cite{hoyer2017xarray}.

Such libraries often mimic the NumPy API, because it lowers the
barrier to entry for newcomers and provides the wider community with a
stable array programming interface. This, in turn, prevents disruptive
schisms like the divergence of Numeric and Numarray.
But exploring new ways of working with arrays is experimental by nature
and, in fact, several promising libraries---such as Theano and Caffe---have
already ceased development. And each time that a user
decides to try a new technology, they must
change import statements and ensure that the new library implements
all the parts of the NumPy API they currently use.

Ideally, operating on specialized arrays using NumPy functions or semantics would
simply work, so that users could write code once, and would then benefit
from switching between NumPy arrays, GPU arrays, distributed arrays,
and so forth, as appropriate.
To support array operations between external array objects, NumPy
therefore added the capability to act as a central coordination
mechanism with a well-specified API (Fig.~\ref{fig:ecosystem}).

To facilitate this \emph{interoperability}, NumPy provides
``protocols'' (or contracts of operation), that allow for specialized arrays to be
passed to NumPy functions (Fig.~\ref{fig:array-protocol}).
NumPy, in turn, dispatches operations to the originating library, as required.
Over four hundred of the most popular
NumPy functions are supported.
The protocols are implemented by widely used libraries such as Dask, CuPy,
xarray, and PyData/Sparse.
Thanks to these developments, users can now, for example, scale
their computation from a single machine to distributed systems using Dask.
The protocols also \textit{compose} well, allowing users to redeploy NumPy
code at scale on distributed, multi-GPU systems via, for instance, CuPy arrays embedded in Dask
arrays. Using NumPy's high-level API, users can leverage highly parallel code
execution on multiple systems with millions of cores, all with minimal code
changes \cite{entschev2019}.

These array protocols are now a key feature of NumPy, and are expected to only
increase in importance.  As with the rest of NumPy, we iteratively refine and
add protocol designs to improve utility and simplify adoption. 

\section*{Discussion}

NumPy combines the expressive power of \emph{array programming},
the performance of C, and
the readability, usability, and versatility of Python in a mature,
well-tested, well-documented, and community-developed library.
Libraries in the scientific Python ecosystem provide fast implementations of most important algorithms.
Where extreme optimization is warranted, compiled languages such as
Cython \cite{behnel2011cython}, Numba \cite{Lam:2015:NLP:2833157.2833162},
and Pythran \cite{guelton2015pythran}, that
extend Python and transparently accelerate bottlenecks, can be
used.
Because of NumPy's simple memory model, it is easy to write low-level, hand-optimized code, usually in C
or Fortran, to manipulate NumPy arrays and pass them back to
Python.
Furthermore, using array protocols, it is possible to utilize the full
spectrum of specialized hardware acceleration with minimal changes to
existing code.

NumPy was initially developed by students, faculty, and researchers to
provide an advanced, open-source array programming library for Python,
which was free to use and unencumbered by license servers, dongles, and the like.
There was a sense of building something consequential together,
for the benefit of many others.  Participating in
such an endeavor, within a welcoming community of like-minded
individuals, held a powerful attraction for many early contributors.

These user-developers frequently had to write code from scratch to solve
their own or their colleagues' problems---often in low-level languages
that precede Python, like Fortran \cite{dongarra2008netlib} and C.
To them, the advantages of an interactive, high-level array library
were evident. The design of this new tool was informed by other
powerful interactive programming languages for scientific computing
such as Basis \cite{dubois1989basis}, Yorick \cite{munro1995using}, R \cite{ihaka1996r},
and APL \cite{iverson1962programming},
as well as commercial languages and environments like IDL and {MATLAB}.

What began as an attempt to add an array object to Python became the
foundation of a vibrant ecosystem of tools.  Now, a large amount of
scientific work depends on NumPy being correct, fast, and stable.  It
is no longer a small community project, but is core scientific
infrastructure.

The developer culture has matured: while initial development was
highly informal, NumPy now has a roadmap and a process for proposing
and discussing large changes.  The project has formal governance
structures and is fiscally sponsored by NumFOCUS, a nonprofit that
promotes open practices in research, data, and scientific computing.
Over the past few years, the project attracted its first funded
development, sponsored by the Moore and Sloan Foundations, and
received an award as part of the Chan Zuckerberg Initiative's
Essentials of Open Source Software program.  With this funding, the
project was (and is) able to have sustained focus over multiple months
to implement substantial new features and improvements.  That said, it
still depends heavily on contributions made by graduate students and
researchers in their free time.

NumPy is no longer \emph{just} the foundational array library underlying
the scientific Python ecosystem, but has also become the standard API
for tensor computation and a central coordinating mechanism between
array types and technologies in Python. Work continues to expand on and
improve these interoperability features.

Over the next decade, we will face several challenges.  New devices will be
developed, and existing specialized hardware will evolve, to meet diminishing
returns on Moore's law.  There will be more, and a wider variety of, data
science practitioners, a significant proportion of whom will be using NumPy.
The scale of scientific data gathering will continue to expand, with the
adoption of devices and instruments such as light sheet microscopes and the
Large Synoptic Survey Telescope (LSST) \cite{jenness2018lsst}.  New generation
languages, interpreters, and compilers, such as Rust
\cite{10.1145/2692956.2663188}, Julia \cite{Julia-2017}, and LLVM
\cite{LLVM:CGO04}, will invent and determine the viability of new concepts and
data structures.

Through various mechanisms described in this paper, NumPy is poised to
embrace such a changing landscape, and to continue playing a leading
role in interactive scientific computation.  To do so will require
sustained funding from government, academia, and industry.  But,
importantly, it will also need a new generation of graduate students
and other developers to engage, to build a NumPy that meets the needs
of the next decade of data science.

\newpage

\section*{Methods}

We use Git for version control and GitHub as the public hosting service for our
official \emph{upstream} repository (\url{https://github.com/numpy/numpy}).
We each work in our own copy (or fork) of the project and use the
upstream repository as our integration point.
To get new code into the upstream repository, we use GitHub's
pull request (PR) mechanism.
This allows us to review code before integrating it as well as to run a
large number of tests on the modified code to ensure that the changes
do not break expected behavior.

We also use GitHub's issue tracking system to collect and triage problems and
proposed improvements.

\subsection*{Library organization}

Broadly, the NumPy library consists of the following parts:
the NumPy array data structure \texttt{ndarray}; the so-called \emph{universal functions};
a set of library functions for manipulating arrays and doing scientific
computation; infrastructure libraries for unit tests and Python package
building; and the program \texttt{f2py} for wrapping Fortran code in Python \cite{peterson2009f2py}.
The \texttt{ndarray} and the universal functions are generally considered
the core of the library.
In the following, we give a brief summary of these components of the
library.

\paragraph{\emph{Core.}}  The \texttt{ndarray} data structure and the
universal functions make up the core of NumPy.

The \texttt{ndarray} is the data structure at the heart of NumPy.
The data structure stores regularly strided homogeneous data types
inside a contiguous block memory, allowing for the efficient representation
of $n$-dimensional data.
More details about the data structure are given in ``The NumPy array:
a structure for efficient numerical computation'' \cite{vanderwalt2011numpy}.

The \emph{universal functions}, or more concisely, \emph{ufuncs},
are functions written in C that implement efficient looping over
NumPy arrays. An important feature of ufuncs is the built-in
implementation of \emph{broadcasting}.  For example, the function
\texttt{arctan2(x, y)} is a ufunc that accepts two values and computes
$\tan^{-1}(y/x)$.  When arrays are passed in as the arguments,
the ufunc will take care of looping over the dimensions of the inputs
in such a way that if, say, \texttt{x} is a 1-D array with length 3, and
\texttt{y} is a 2-D array with shape $2 \times 1$, the output will be
an array with shape $2 \times 3$ (Fig.~\ref{fig:array-concepts}c).
The ufunc machinery takes care
of calling the function with all the appropriate combinations of
input array elements to complete the output array.
The elementary arithmetic operations of addition, multiplication, etc.,
are implemented as ufuncs, so that broadcasting also applies to expressions
such as \texttt{x + y * z}.

\paragraph{\emph{Computing libraries.}}
NumPy provides a large library of functions for array manipulation
and scientific computing, including functions for: creating, reshaping,
concatenating, and padding arrays; searching, sorting and counting data
in arrays; computing elementary statistics, such as the mean, median,
variance, and standard deviation; file I/O; and more.

A suite of functions for computing the \emph{fast Fourier transform (FFT)}
and its inverse is provided.

NumPy's linear algebra library includes functions for: solving linear
systems of equations; computing various functions of a matrix, including
the determinant, the norm, the inverse, and the pseudo-inverse;
computing the Cholesky, eigenvalue, and singular value decompositions of a matrix;
and more.

The random number generator library in NumPy provides alternative
\emph{bit stream generators} that provide the core function of generating
random integers.
A higher-level generator class that implements an assortment of
probability distributions is provided. It includes the beta, gamma
and Weibull distributions, the univariate and multivariate normal
distributions, and more.

\paragraph{\emph{Infrastructure libraries.}} NumPy provides utilities
for writing tests and for building Python packages.

The \texttt{testing} subpackage provides functions such as
\texttt{assert\_allclose(actual, desired)} that may be used in
test suites for code that uses NumPy arrays.

NumPy provides the subpackage \texttt{distutils} which includes functions and classes
to facilitate configuration, installation, and packaging of libraries depending on NumPy.
These can be used, for example, when publishing to the PyPI website.

\paragraph{\emph{F2PY.}}  The program \texttt{f2py} is a tool for
building NumPy-aware Python wrappers of Fortran functions.
NumPy itself does not use any Fortran code;  F2PY is part of NumPy
for historical reasons.

\subsection*{Governance}

NumPy adopted an official Governance Document on October~5,
2015 \cite{NumPyProjectGovernance}.
Project decisions are usually made by consensus of interested contributors.
This means that, for most decisions, everyone is entrusted with veto power.
A Steering Council, currently composed of 12~members, facilitates this
process and oversees daily development of the project by contributing code
and reviewing contributions from the community.

NumPy's official Code of Conduct was approved on September~1, 2018 \cite{NumPyCodeofConduct}.
In brief, we strive to:
\emph{be open};
\emph{be empathetic, welcoming, friendly, and patient};
\emph{be collaborative};
\emph{be inquisitive}; and
\emph{be careful in the words that we choose}.
The Code of Conduct also specifies how breaches can be reported and outlines
the process for responding to such reports.

\subsection*{Funding}

In 2017, NumPy received its first large grants totaling 1.3M USD from the
Gordon \& Betty Moore and the Alfred P. Sloan foundations.
Stéfan van der Walt is the PI and manages four programmers working on the project.
These two grants focus on addressing the technical debt accrued over the years and
on setting in place standards and architecture to encourage more sustainable development.

NumPy received a third grant for 195K USD from the Chan Zuckerberg
Initiative at the end of 2019 with Ralf Gommers as the PI.
This grant focuses on better serving NumPy's large number of beginning
to intermediate level users and on growing the community of NumPy
contributors.
It will also provide support to OpenBLAS, on which NumPy depends for
accelerated linear algebra.

Finally, since May 2019 the project receives a small amount annually from
Tidelift, which is used to fund things like documentation and website
improvements.

\subsection*{Developers}

NumPy is currently maintained by a group of 23 contributors with commit rights
to the NumPy code base. Out of these, 17 maintainers were active in
2019, 4 of whom were paid to work on the project full-time.
Additionally, there are a few long term developers who contributed and maintain
specific parts of NumPy, but are not officially maintainers.

Over the course of its history, NumPy has attracted PRs by 823 contributors.
However, its development relies heavily on a small number
of active maintainers, who share more than half of the contributions among
themselves.

\begin{figure}
    \centering
    \includegraphics[width=0.9\linewidth]{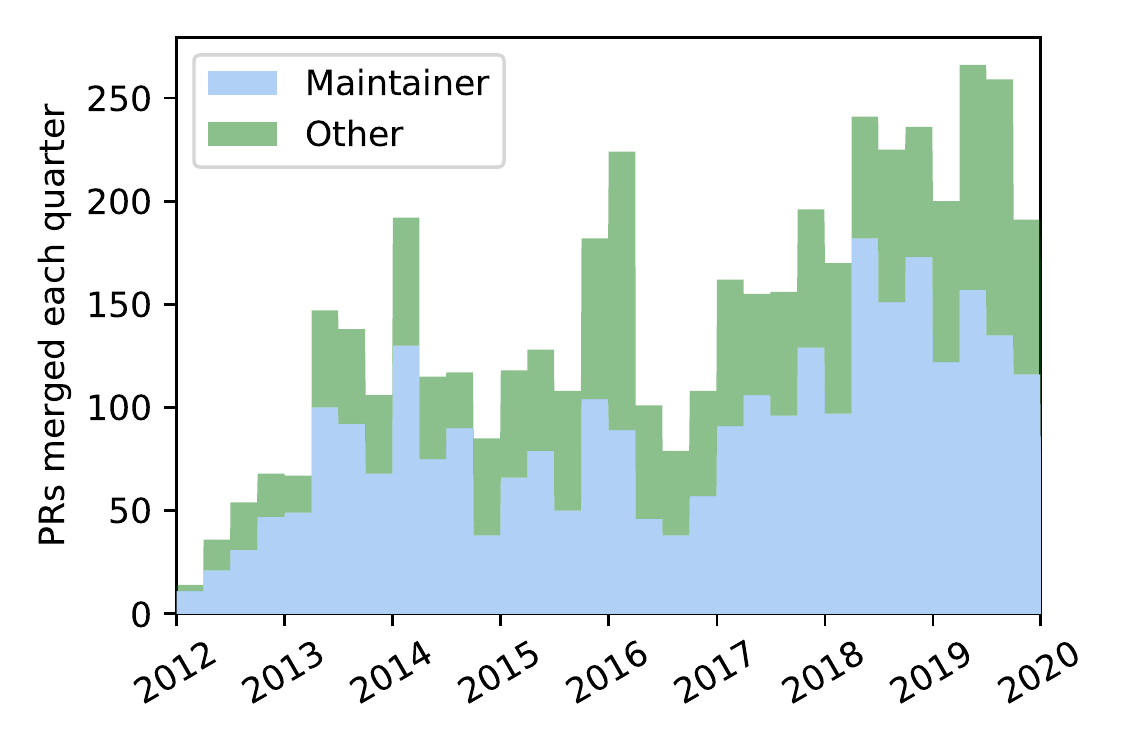}
    \caption{\textbf{Number of pull requests merged into the NumPy master branch for each
        quarter since 2012.} The total number of PRs is indicated with the
        lower blue area showing the portion contributed by current or previous
        maintainers.}\label{fig:prs-over-time}
\end{figure}

At a release cycle of about every half year, the five recent releases in the years
2018 and 2019 have averaged about 450~PRs each,\footnote{
    Note that before mid 2011, NumPy development did not happen on \url{github.com}.
    All data provided here is based on the development which happened through GitHub
    PRs. In some cases contributions by maintainers may not be categorized as such.}
with each release attracting more than a hundred new contributors.
Figure~\ref{fig:prs-over-time} shows the number of PRs merged into the NumPy
master branch.
Although the number of PRs being merged fluctuates,
the plot indicates an increased number of contributions over the past
years.

\subsection*{Community calls}

The massive number of scientific Python packages that
built on NumPy meant that it had an unusually high need for stability.
So to guide our development we formalized the feature proposal process, and
constructed a development roadmap with extensive input and feedback from the
community.

Weekly community calls alternate between triage and
higher level discussion.  The calls not only involve developers from
the community, but provide a venue for vendors and other external
groups to provide input.  For example, after Intel produced a forked
version of NumPy, one of their developers joined a call to discuss
community concerns.

\subsection*{NumPy enhancement proposals}

Given the complexity of the codebase and the massive number of projects depending
on it, large changes require careful planning and substantial work.
NumPy Enhancement Proposals (NEPs) are modeled after
Python Enhancement Proposals (PEPs) for ``proposing major new
features, for collecting community input on an issue, and for
documenting the design decisions that have gone into
Python''\footnote{\url{https://numpy.org/neps/nep-0000.html}}.
Since then there have been 19 proposed NEPS---6 have been implemented,
4 have been accepted and are being implemented, 4 are under
consideration, 3 have been deferred or superseded, and 2 have been rejected
or withdrawn.

\subsection*{Central role}

NumPy plays a central role in building and standardizing much of the scientific
Python community infrastructure.
NumPy's docstring standard is now widely adopted.
We are also now using the NEP system as a way to help coordinate the larger
scientific Python community.
For example, in NEP 29, we recommend, along with leaders from various other
projects, that all projects across the Scientific Python ecosystem adopt a
common ``time window-based'' policy for support of Python and NumPy versions.
This standard will simplify downstream project and release planning.

\subsection*{Wheels build system}

A Python \emph{wheel} \cite{PEP427} is a standard file format for
distributing Python libraries.  In addition to Python code, a
wheel may include compiled C extensions and other binary data.
This is important, because many libraries, including NumPy,
require a C compiler and other build tools to build the software
from the source code, making it difficult for many users to install
the software on their own.  The introduction of wheels to the Python
packaging system has made it much easier for users to install
precompiled libraries.

A GitHub repository containing scripts to build NumPy wheels has
been configured so that a simple commit to the repository triggers
an automated build system that creates NumPy wheels for several
computer platforms, including Windows, Mac OSX and Linux.  The wheels
are uploaded to a public server and made available for anyone to use.
This system makes it easy for users to install precompiled versions
of NumPy on these platforms.

The technology that is used to build the wheels evolves continually.
At the time this paper is being written, a key component is the
\texttt{multibuild} suite of tools developed by Matthew Brett and
other developers \cite{multibuild}.  Currently, scripts using
\texttt{multibuild} are written for the continuous integration
platforms Travis-CI (for Linux and Mac OSX) and Appveyor
(for Windows).

\subsection*{Recent technical improvements}

With the recent infusion of funding and a clear process for coordinating with
the developer community, we have been able to tackle a number of important
large scale changes.
We highlight two of those below, as well as changes made to our testing
infrastructure to support hardware platforms used in large scale computing.

\subsection*{Array function protocol}

A vast number of projects are built on NumPy;
these projects are consumers of the NumPy API.
Over the last several years, a growing number of projects are providers of
a \emph{NumPy-like API} and array objects targeting audiences with specialized
needs beyond NumPy's capabilities.
For example, the NumPy API is implemented by several popular tensor computation
libraries including CuPy\footnote{\url{https://cupy.chainer.org/}},
JAX\footnote{\url{https://jax.readthedocs.io/en/latest/jax.numpy.html}},
and Apache MXNet\footnote{\url{https://numpy.mxnet.io/}}.
PyTorch\footnote{\url{https://pytorch.org/tutorials/beginner/blitz/tensor\_tutorial.html}}
and Tensorflow\footnote{\url{https://www.tensorflow.org/tutorials/customization/basics}}
provide tensor APIs with NumPy-inspired semantics.
It is also implemented in packages that support sparse arrays
such as \texttt{scipy.sparse} and PyData/Sparse.
Another notable example is Dask, a library for parallel computing in
Python.  Dask adopts the NumPy API and therefore presents a familiar
interface to existing NumPy users, while adding powerful abilities to
parallelize and distribute tasks.

The multitude of specialized projects creates the difficulty that consumers
of these NumPy-like APIs write code specific to a single project and do not support
all of the above array providers.
This is a burden for users relying on the specialized array-like, since
a tool they need may not work for them.
It also creates challenges for end-users who need to transition
from NumPy to a more specialized array.
The growing multitude of specialized projects with NumPy-like APIs threatened
to again fracture the scientific Python community.

To address these issues NumPy has the goal of providing the fundamental
API for \emph{interoperability} between the various NumPy-like APIs.
An earlier step in this direction was the implementation of the
\texttt{\_\_array\_ufunc\_\_} protocol in NumPy 1.13, which enabled interoperability
for most mathematical functions \cite{NEP13}.
In 2019 this was expanded more generally with the inclusion of the
\texttt{\_\_array\_function\_\_} protocol into NumPy~1.17.
These two protocols allow providers of array objects to be interoperable
with the NumPy API: their arrays work correctly with almost all NumPy functions \cite{NEP18}.
For the users relying on specialized array projects it means that even though
much code is written specifically for NumPy arrays and uses the NumPy API as
\texttt{import numpy as np}, it can nevertheless work for them.
For example, here is how a CuPy GPU array can be passed through NumPy for
processing, with all operations being dispatched back to CuPy:

\begin{lstlisting}
import numpy as np
import cupy as cp

x_gpu = cp.array([1, 2, 3])
y = np.sum(x_gpu)  # Returns a GPU array
\end{lstlisting}

Similarly, user defined functions composed using NumPy can now be
applied to, e.g., multi-node distributed Dask arrays:

\begin{lstlisting}
import numpy as np
import dask.array as da


def f(x):
    """Function using NumPy API calls"""
    y = np.tensordot(x, x.T)
    return np.mean(np.log(y + 1))


x_local = np.random.random([10000, 10000])  # random local array
x_distr = da.random.random([10000, 10000])  # random distributed array

f(x_local)  # returns a NumPy array
f(x_distr)  # works, returns a Dask array
\end{lstlisting}

\subsection*{Random number generation}

The NumPy \texttt{random} module provides pseudorandom numbers from a wide range of
distributions. In legacy versions of NumPy, simulated random values are produced
by a \texttt{RandomState} object that: handles seeding and state initialization;
wraps the core pseudorandom number generator based on a Mersenne Twister
implementation\footnote{to be precise, the standard 32-bit version of MT19937};
interfaces with the underlying code that transforms random bits into
variates from other distributions; and supplies a singleton instance exposed in
the root of the random module.

The \texttt{RandomState} object makes a compatibility guarantee so that a fixed
seed and sequence of function calls produce the same set of values. This
guarantee has slowed progress since improving the underlying code requires
extending the API with additional keyword arguments. This guarantee continues to
apply to \texttt{RandomState}.

NumPy 1.17 introduced a new API for generating random numbers that use a more
flexible structure that can be extended by libraries or end-users. The new API
is built using components that separate the steps required to generate random
variates. Pseudorandom bits are generated by a bit generator. These bits are
then transformed into variates from complex distributions by a generator.
Finally, seeding is handled by an object that produces sequences of high-quality
initial values.

Bit generators are simple classes that manage the state of an underlying
pseudorandom number generator. NumPy ships with four bit generators. The default
bit generator is a 64-bit implementation of the Permuted Congruential Generator
\cite{pcg64} (\texttt{PCG64}). The three other bit generators are a 64-bit version
of the Philox generator \cite{random123} (\texttt{Philox}), Chris Doty-Humphrey's
Small Fast Chaotic generator \cite{practrand} (\texttt{SFC64}), and the 32-bit
Mersenne Twister \cite{mt19937} (\texttt{MT19937}) which has been used in older
versions of NumPy.\footnote{The
\href{https://github.com/bashtage/randomgen}{randomgen project} supplies a wide
range of alternative bit generators such as a cryptographic counter-based
generators (\texttt{AESCtr}) and generators that expose hardware random number
generators (\texttt{RDRAND}) \cite{randomgen}.} Bit generators provide
functions, exposed both in Python and C, for generating random integer
and floating point numbers.

The \texttt{Generator} consumes one of the bit generators and produces variates
from complicated distributions. Many improved methods for generating random
variates from common distributions were implemented, including the Ziggurat
method for normal, exponential and gamma variates \cite{ziggurat}, and Lemire's
method for bounded random integer generation \cite{lemire}. The \texttt{Generator}
is more similar to the legacy \texttt{RandomState}, and its API is substantially
the same. The key differences all relate to state management, which has been
delegated to the bit generator. The \texttt{Generator} does not make the same
stream guarantee as the \texttt{RandomState} object, and so variates may differ
across versions as improved generation algorithms are
introduced.\footnote{Despite the removal of the compatibility guarantee, simple
reproducibility across versions is encouraged, and minor changes that do not
produce meaningful performance gains or fix underlying bug are not generally
adopted.}

Finally, a \texttt{SeedSequence} is used to initialize a bit generator. The seed
sequence can be initialized with no arguments, in which case it reads entropy
from a system-dependent provider, or with a user-provided seed. The seed
sequence then transforms the initial set of entropy into a sequence of
high-quality pseudorandom integers, which can be used to initialize multiple bit
generators deterministically. The key feature of a seed sequence is that
it can be used to spawn child \texttt{SeedSequence}s to initialize
multiple distinct bit generators.
This capability allows a seed sequence to facilitate large distributed applications
where the number of workers required is not known. The sequences generated from
the same initial entropy and spawns are fully deterministic to ensure
reproducibility.

The three components are combined to construct a complete random number
generator.

\begin{lstlisting}
from numpy.random import (
    Generator,
    PCG64,
    SeedSequence,
)

seq = SeedSequence(1030424547444117993331016959)
pcg = PCG64(seq)
gen = Generator(pcg)
\end{lstlisting}

This approach retains access to the seed sequence which can then be
used to spawn additional generators.

\begin{lstlisting}
children = seq.spawn(2)
gen_0 = Generator(PCG64(children[0]))
gen_1 = Generator(PCG64(children[1]))
\end{lstlisting}

While this approach retains complete flexibility, the method
\texttt{np.random.default\_rng} can be used to instantiate a \texttt{Generator} when
reproducibility is not needed.

The final goal of the new API is to improve extensibility. \texttt{RandomState} is
a monolithic object that obscures all of the underlying state and functions. The
component architecture is one part of the extensibility improvements. The
underlying functions (written in C) which transform the output of a bit
generator to other distributions are available for use in CFFI. This allows the
same code to be run in both NumPy and dependent that can consume CFFI, e.g.,
Numba. Both the bit generators and the low-level functions can also be used in C
or Cython code.\footnote{As of 1.18.0, this scenario requires access to the
NumPy source. Alternative approaches that avoid this extra step are being
explored.}

\subsection*{Testing on multiple architectures}

At the time of writing the two fastest supercomputers in the
world, Summit and Sierra, both have IBM POWER9 architectures
\cite{top500nov2019}. In late 2018, Astra, the first ARM-based
supercomputer to enter the TOP500 list, went into production
\cite{astra-wiki}. Furthermore, over 100 billion ARM processors have been
produced as of 2017 \cite{arm-architecture}, making it the most 
widely used instruction set architecture in the world.

Clearly there are motivations for a large scientific computing
software library to support POWER and ARM architectures. We've extended
our continuous integration (CI) testing to include \texttt{ppc64le}
(POWER8 on Travis CI) and ARMv8 (on Shippable service). We also test
with the s390x architecture (IBM Z CPUs on Travis CI) so that we
can probe the behavior of our library on a big-endian machine.
This satisfies one of the major components of
improved CI testing laid out in a version of our roadmap
\cite{numpy-roadmap}---specifically, ``CI for more exotic
platforms."

PEP 599 \cite{PEP599} lays out a plan for new Python binary wheel
distribution support, \texttt{manylinux2014}, that adds
support for a number of architectures supported by the CentOS
Alternative Architecture Special Interest Group, including
ARMv8, ppc64le, as well as s390x. We are thus well-positioned
for a future where provision of binaries on these architectures
will be expected for a library at the base of the ecosystem.
 
\section*{Acknowledgments}

We thank Ross Barnowski, Paul Dubois, Michael Eickenberg, and Perry Greenfield, who
suggested text and provided helpful feedback on the manuscript.

We also thank the many members of the community who provided
feedback, submitted bug reports, made improvements to the documentation,
code, or website, promoted NumPy's use in their scientific fields, and built
the vast ecosystem of tools and libraries around NumPy.
We also gratefully acknowledge the Numeric and Numarray developers
on whose work we built.  

Jim Hugunin wrote Numeric in 1995, while a graduate student at MIT.
Hugunin based his package on previous work by Jim Fulton, then working at the
US Geological Survey, with input from many others.
After he graduated, Paul Dubois at the Lawrence Livermore National Laboratory
became the maintainer.
Many people contributed to the project including T.E.O. (a co-author
of this paper), David Ascher, Tim Peters, and Konrad Hinsen.

In 1998 the Space Telescope Science Institute started using Python
and in 2000 began developing a new array package called Numarray, written
almost entirely by Jay Todd Miller, starting from a prototype developed by
Perry Greenfield.  Other contributors included Richard L. White, J. C. Hsu,
Jochen Krupper, and Phil Hodge.
The Numeric/Numarray split divided the community, yet ultimately pushed
progress much further and faster than would otherwise have been possible. 

Shortly after Numarray development started, T.E.O. took over maintenance of
Numeric. In 2005, he led the effort and did most of the work to unify Numeric
and Numarray, and produce the first version of NumPy.

Eric Jones co-founded (along with T.E.O. and P.P.) the SciPy community, gave early feedback on array
implementations, and provided funding and travel support to several
community members.
Numerous people contributed to the creation and
growth of the larger SciPy ecosystem, which gives NumPy much of its
value. Others injected new energy and ideas by creating experimental
array packages.

K.J.M. and S.J.v.d.W. were funded in part by the Gordon and Betty Moore
Foundation through Grant GBMF3834 and by the Alfred P. Sloan Foundation through
Grant 2013-10-27 to the University of California, Berkeley.
S.J.v.d.W., S.B., M.P., and W.W. were funded in part by the Gordon
and Betty Moore Foundation through Grant GBMF5447 and by the Alfred
P. Sloan Foundation through Grant G-2017-9960 to the University of
California, Berkeley.

\section*{Author Contributions Statement}

K.J.M. and S.J.v.d.W. composed the manuscript with input from others.
S.B., R.G., K.S., W.W., M.B., and T.J.R. contributed text.
All authors have contributed significant code, documentation, and/or expertise
to the NumPy project.
All authors reviewed the manuscript.

\section*{Competing Interests}

The authors declare no competing interests.

\end{document}